\documentclass[aps,preprint,showpacs,groupedaddress]{revtex4-1}
\usepackage{amssymb}
\usepackage{mathrsfs}
\usepackage{amsmath}
\usepackage{mathtools}
\usepackage{pstricks}
\usepackage{color}
\usepackage{graphicx}
\usepackage{amsthm}

\begin{document}

%\preprint{}

\title{\bf Diffeomorphism symmetry effect on the O($N$) universality class}



\author{H. A. S. Costa}
\email{hascosta@ufpi.edu.br}
\affiliation{\it Departamento de F\'\i sica, Universidade Federal do Piau\'\i, 64049-550, Teresina, PI, Brazil}
%\affiliation{\it Escola Polit\'{e}cnica de Pernambuco, Universidade de Pernambuco, 50720-001, Recife, PE, Brazil}
%\affiliation{\it Instituto de F\'{i}sica, Universidade Federal de Alagoas, 57072-900, Macei\'{o}, AL, Brazil}

\author{P. R. S. Carvalho}
\email{prscarvalho@ufpi.edu.br}
\affiliation{\it Departamento de F\'\i sica, Universidade Federal do Piau\'\i, 64049-550, Teresina, PI, Brazil}

%\thanks{}


%\date{\today}

\begin{abstract}
We probe the effect of diffeomorphism symmetry on the critical exponents values for massive O($N$) $\lambda\phi^{4}$ scalar field theories in curved spacetime. We apply field-theoretic renormalization group tools, where we use only momentum coordinates as opposed to the earlier standard procedure with both space and momentum coordinates, for computing analytically the critical exponents up to three-loop order for one of them and up to two-loop level for the remaining ones. We found that the curved spacetime critical exponents are the same as their flat spacetime counterparts, at least up to the level considered. At the conclusions we present the physical interpretation for the results.  
\end{abstract}

%\keywords{}

\maketitle


\section{Introduction}

The diffeomorphism symmetry is one of the most important symmetries in describing the behavior of physical systems. It is essential, for example, in constructing the general theory of relativity that successfully describes one of the four fundamental interactions of nature, namely gravitation \cite{Misner.Thorne.Wheeler,Wald}. So any theory involving the gravitational field must be invariant under general coordinate transformations. In the theory approached here, a fluctuating self-interacting scalar quantum field $\phi$ interacts with gravity through a general curved background $R$. The critical properties of the theory are analog to that of a system undergoing a continuous phase transition characterized by a set of critical exponents. For example, a magnetic system whose mean value of the quantum field would be identified to the magnetization of the system. The mass of the quantum field would be proportional to the difference between some arbitrary temperature $T$ and the critical one $T_{c}$, namely $m^{2}\propto T - T_{c}$. These critical properties are a result of the fluctuations of the quantum field when these fluctuations are taken into account at the many length scales present in the system thus giving rise to the radiative quantum corrections to the critical exponents \cite{PhysRevB.4.3174,PhysRevB.4.3184}. These corrections are called corrections to mean field theory since in the mean field theory or Landau approximation \cite{Stanley} the fluctuations are neglected. On the technical viewpoint, these corrections are perturbative corrections when the interacting theory represented by a Lagrangian density is perturbatively expanded in the coupling constant of the self-interacting quantum field. This Lagrangian density must be constructed by satisfying certain constraints. The first of them is that all odd powers of the quantum field are forbidden, because in a system undergoing a continuous phase transition the total energy of the system remains invariant when there is a simultaneous discrete change of all spin orientations thus characterizing the symmetry $\phi \rightarrow -\phi$. The second one is that only operators with canonical dimension smaller than or equal to four, called relevant and marginal operators, respectively, must remain, since that ones whose canonical dimensions are greater than four do not contribute to the critical properties of the system \cite{Amit}. These two constraints are enough to construct the Lagrangian density where the quantum field is embedded in a flat spacetime. Additionally, in a curved spacetime, the Lagrangian density must be integrated over an invariant volume, defined on a Riemannian manifold and now must be invariant under diffeomorphism transformations. Thus we have to introduce the overall factor $\sqrt{g}$, where $g$ is given by $g = det(g^{\mu\nu})$, representing a minimal coupling between the quantum field and gravity. Now, the most general allowed interaction satisfying the constraints aforementioned is a non-minimal one of the form $\xi R \phi^{2}$ \cite{Birrell.Davies,Vilenkin.Ford,Brown.Collins,Hathrell,Birrell,Bunch.Panangaden.Parker,Bunch.Panangaden,Toms,
Bunch.Parker,Panangaden}, where $\xi$ is the non-minimal coupling constant. Once the Lagrangian density has been defined, we are plagued with the following problem: the perturbative expansion is divergent and if we want a reliable theory, we have to get rid these divergences. This task is handled by the renormalization group technique \cite{Wilson197475}. These divergences come from the fact of treating the nontrivial problem of many degrees of freedom interacting with each other at many length scales. Thus, when two quantum fields interact at the same point of spacetime, these divergences emerge \cite{Wilsonmanylength}. In our analog example, these divergences would be the interaction of magnetization domains of many sizes \cite{Huang}. In fact, in the field theoretic framework approached here the ultraviolet divergences of the massive theory are present in the one-particle-irreducible ($1$PI) vertex parts $\Gamma^{(n)}$. They contain all information about the scaling properties of the theory, since the thermodynamic functions near continuous phase transitions display a simple scale behavior. From all $1$PI vertex parts, we need to renormalize only two of them, namely $\Gamma^{(2)}$ and $\Gamma^{(4)}$, called primitively divergent $1$PI vertex parts. The others $1$PI vertex parts, for $n > 4$, get automatically renormalized if the primitively divergent ones are renormalized. This result is a consequence of the skeleton expansion \cite{ZinnJustin}. From the scaling properties of $\Gamma^{(2)}$ and $\Gamma^{(4)}$ we can obtain the four scaling relations among the six critical exponents \cite{Amit}, thus being needed to evaluate independently only two of them, $\eta$ and $\nu$ for example. An interesting feature of systems undergoing a continuous phase transition is that completely different systems as a fluid and a ferromagnet share the same set of critical exponents. When this happens, we say that they belong to the same universality class. This occurs when they have the same dimension $d$, $N$ and symmetry of some $N$-component order parameter (the magnetization for magnetic systems) if the interactions of their constituents are of short- or long-range type. The universality class treated here will be the O($N$) one, which encompasses the specific models: Ising ($N = 1$), XY ($N = 2$), Heisenberg ($N = 3$), self-avoiding random walk ($N = 0$), spherical ($N \rightarrow \infty$) etc. \cite{Pelissetto2002549}. Of course, the $d$ \cite{PhysRevB.86.155112,PhysRevE.71.046112} and $N$ \cite{PhysRevLett.110.141601,Butti2005527,PhysRevB.54.7177} parameters are easier to probe than symmetry effects \cite{CARVALHO2017290,Carvalho2017}. Probing the effect of the latter in a curved spacetime is the aim of the present work. 

\par For renormalizing the theory approached here we will apply the Bogoliubov-Parasyuk-Hepp-Zimmermann (BPHZ) renormalization scheme \cite{BogoliubovParasyuk,Hepp,Zimmermann}. In this renormalization scheme, we begin with the primitively divergent $1$PI vertex parts expanded up to one-loop order. At this order, the $1$PI vertex parts are divergent. Thus we add to them terms such that the primitively divergent $1$PI vertex parts turn out to be finite. These terms are called counterterms and can be viewed as being generated by added terms to the initially divergent Lagrangian density. Now the theory get renormalized at this loop order. We now proceed to the two-loop level and repeat the same steps for obtaining a renormalized theory up to two-loop order. We can repeat this procedure up to a given desired order such that we always attain a renormalized theory at that level. Then we can start from the renormalized Lagrangian density at a given loop order. This will be done up to next-to-leading order.

\section{Next-to-leading order bare theory and its renormalization}\label{Next-to-leading order bare theory and its renormalization} 

\par The theory approached here is described by the bare Lagrangian density
\begin{eqnarray}\label{bare Lagrangian density}
\mathcal{L}_{B} =  \sqrt{g}\frac{1}{2}\left(\partial_{\mu}\phi_{B}\partial^{\mu}\phi_{B} + m_{B}^{2}\phi_{B}^{2} + \xi_{B} R\phi_{B}^{2}\right) + \sqrt{g}\frac{\lambda_{B}}{4!}\phi_{B}^{4}
\end{eqnarray}
embedded on a curved spacetime with a Riemannian metric signature on $d$ dimensions. We have to renormalize the divergences of the theory through the renormalization constants \cite{BogoliubovParasyuk,Hepp,Zimmermann}
%\begin{eqnarray}\label{gtfrdrdes}
%&&\Gamma^{(2)}_{B} = \parbox{12mm}{\includegraphics[scale=1.0]{fig9.eps}}^{-1} - \frac{1}{2}\hspace{1mm}\parbox{12mm}{\includegraphics[scale=1.0]{fig1.eps}} - \frac{1}{4}\hspace{1mm}\parbox{12mm}{\includegraphics[scale=1.0]{fig2.eps}} - \frac{1}{6}\hspace{1mm}\parbox{12mm}{\includegraphics[scale=1.0]{fig6.eps}} \nonumber \\ && - \frac{1}{4}\hspace{1mm}\parbox{10mm}{\includegraphics[scale=0.8]{fig7.eps}} - \frac{1}{8}\hspace{1mm}\parbox{10mm}{\includegraphics[scale=0.8]{fig3.eps}} - \frac{1}{8}\hspace{1mm}\parbox{10mm}{\includegraphics[scale=0.8]{fig4.eps}} - \frac{1}{4}\hspace{1mm}\parbox{10mm}{\includegraphics[scale=0.8]{fig8.eps}}  \nonumber \\&& -\frac{1}{12}\hspace{1mm}\parbox{10mm}{\includegraphics[scale=0.8]{fig5.eps}}, 
%\end{eqnarray}
%\begin{eqnarray}
%&&\Gamma^{(4)}_{B} = -\parbox{10mm}{\includegraphics[scale=0.09]{fig29.eps}} - \frac{1}{2}\hspace{1mm}\parbox{10mm}{\includegraphics[scale=1.0]{fig10.eps}} + 2 \hspace{1mm} perm.   \nonumber \\ && - \frac{1}{4}\hspace{1mm}\parbox{16mm}{\includegraphics[scale=1.0]{fig11.eps}} + 2 \hspace{1mm} perm.  -  \frac{1}{2}\hspace{1mm}\parbox{12mm}{\includegraphics[scale=0.8]{fig21.eps}} + 5 \hspace{1mm} perm. \nonumber \\&& -  \frac{1}{2}\hspace{1mm}\parbox{12mm}{\includegraphics[scale=1.0]{fig13.eps}} + 2 \hspace{1mm} perm.,
%\end{eqnarray} 
\begin{eqnarray}
\phi = Z_{\phi}^{-1/2}\phi_{B}, 
\end{eqnarray}
\begin{eqnarray}
u = \mu^{-\epsilon}Z_{\phi}^{2}\lambda_{B}/Z_{u}, 
\end{eqnarray}
\begin{eqnarray}
m^{2} = Z_{\phi}m_{B}^{2}/Z_{m^{2}},
\end{eqnarray}
\begin{eqnarray}
\xi = Z_{\phi}\xi_{B}/Z_{\xi},
\end{eqnarray}
where
\begin{eqnarray}\label{Zphi}
&& Z_{\phi}(u,\epsilon^{-1}) = 1 + \frac{1}{P^{2}} \Biggl[ \frac{1}{6} \mathcal{K} 
\left(\parbox{12mm}{\includegraphics[scale=1.0]{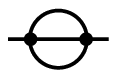}}
\right) \Biggr|_{m^2 = R = 0} S_{\parbox{8mm}{\includegraphics[scale=0.5]{fig6.eps}}} + \frac{1}{4} \mathcal{K} 
\left(\parbox{12mm}{\includegraphics[scale=1.0]{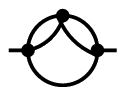}} \right) \Biggr|_{m^2 = R = R_{\mu\nu} = 0} S_{\parbox{6mm}{\includegraphics[scale=0.5]{fig7.eps}}} +  \nonumber \\&& \frac{1}{3} \mathcal{K}
  \left(\parbox{12mm}{\includegraphics[scale=1.0]{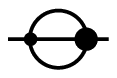}} \right) S_{\parbox{6mm}{\includegraphics[scale=0.5]{fig26.eps}}} \Biggr], 
\end{eqnarray}

\begin{eqnarray}\label{Zg}
&&Z_{u}(u,\epsilon^{-1}) = 1 +  \frac{1}{\mu^{\epsilon}u} \Biggl[ \frac{1}{2} \mathcal{K} 
\left(\parbox{10mm}{\includegraphics[scale=1.0]{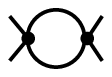}} + 2 \hspace{1mm} perm.
\right) S_{\parbox{8mm}{\includegraphics[scale=0.5]{fig10.eps}}} +  \frac{1}{4} \mathcal{K} 
\left(\parbox{17mm}{\includegraphics[scale=1.0]{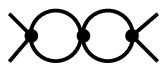}} + 2 \hspace{1mm} perm. \right) S_{\parbox{10mm}{\includegraphics[scale=0.5]{fig11.eps}}} \nonumber \\&&  +  \frac{1}{2} \mathcal{K} 
\left(\parbox{12mm}{\includegraphics[scale=.8]{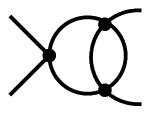}} + 5 \hspace{1mm} perm. \right) S_{\parbox{10mm}{\includegraphics[scale=0.4]{fig21.eps}}} +  \frac{1}{2} \mathcal{K} 
\left(\parbox{10mm}{\includegraphics[scale=1.0]{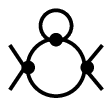}} + 2 \hspace{1mm} perm.
\right) S_{\parbox{10mm}{\includegraphics[scale=0.5]{fig13.eps}}} +  \nonumber \\&& \mathcal{K}
  \left(\parbox{10mm}{\includegraphics[scale=1.0]{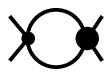}} + 2 \hspace{1mm} perm. \right) S_{\parbox{6mm}{\includegraphics[scale=0.5]{fig25.eps}}} + \mathcal{K}
  \left(\parbox{10mm}{\includegraphics[scale=1.0]{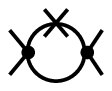}} + 2 \hspace{1mm} perm. \right) S_{\parbox{6mm}{\includegraphics[scale=0.5]{fig24.eps}}} \Biggr],
\end{eqnarray}

\begin{eqnarray}\label{Zm2}
&& Z_{m^{2}}(u,\epsilon^{-1}) = 1 + \frac{1}{m^{2}} \Biggl[ \frac{1}{2} \mathcal{K} 
\left(\parbox{12mm}{\includegraphics[scale=1.0]{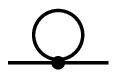}}
\right)\Biggr|_{R = 0} S_{\parbox{10mm}{\includegraphics[scale=0.5]{fig1.eps}}} +  \frac{1}{4} \mathcal{K} 
\left(\parbox{12mm}{\includegraphics[scale=1.0]{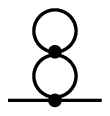}} \right)\Biggr|_{R = 0} S_{\parbox{6mm}{\includegraphics[scale=0.5]{fig2.eps}}} +  \nonumber \\&& \frac{1}{2} \mathcal{K}
  \left(\parbox{8mm}{\includegraphics[scale=1.0]{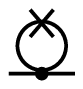}} \right)\Biggr|_{R = 0} S_{\parbox{6mm}{\includegraphics[scale=0.5]{fig22.eps}}} +  \frac{1}{2} \mathcal{K}
  \left(\parbox{12mm}{\includegraphics[scale=1.0]{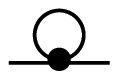}} \right)\Biggr|_{R = 0} S_{\parbox{6mm}{\includegraphics[scale=0.5]{fig23.eps}}} + \frac{1}{6} \mathcal{K}
  \left(\parbox{12mm}{\includegraphics[scale=1.0]{fig6.eps}} \right)\Biggr|_{P^2 = R = 0} S_{\parbox{6mm}{\includegraphics[scale=0.5]{fig6.eps}}} \Biggr], 
\end{eqnarray}

\begin{eqnarray}\label{Zxi}
&& Z_{\xi}(u,\epsilon^{-1}) = 1 + \frac{1}{R} \Biggl[ \frac{1}{2} \mathcal{K} 
\left(\parbox{12mm}{\includegraphics[scale=1.0]{fig1.eps}}
\right)\Biggr|_{m^{2} = 0} S_{\parbox{10mm}{\includegraphics[scale=0.5]{fig1.eps}}} +  \frac{1}{4} \mathcal{K} 
\left(\parbox{12mm}{\includegraphics[scale=1.0]{fig2.eps}} \right)\Biggr|_{m^{2} = 0} S_{\parbox{6mm}{\includegraphics[scale=0.5]{fig2.eps}}} +  \nonumber \\&& \frac{1}{2} \mathcal{K}
  \left(\parbox{8mm}{\includegraphics[scale=1.0]{fig22.eps}} \right)\Biggr|_{m^{2} = 0} S_{\parbox{6mm}{\includegraphics[scale=0.5]{fig22.eps}}} +  \frac{1}{2} \mathcal{K}
  \left(\parbox{12mm}{\includegraphics[scale=1.0]{fig23.eps}} \right)\Biggr|_{m^{2} = 0} S_{\parbox{6mm}{\includegraphics[scale=0.5]{fig23.eps}}} +  \frac{1}{6} \mathcal{K}
  \left(\parbox{12mm}{\includegraphics[scale=1.0]{fig6.eps}} \right)\Biggr|_{P^{2} = m^{2} = 0} S_{\parbox{6mm}{\includegraphics[scale=0.5]{fig6.eps}}} \Biggr] 
\end{eqnarray}
whose renormalized $1$PI vertex parts satisfy the Callan-Symanzik equation
\begin{eqnarray}
\Bigg[\mu\frac{\partial}{\partial\mu} + \beta(u)\,\frac{\partial}{\partial u} - n\,\gamma(u) + \gamma_{m^{2}}(u)\,m^{2}\frac{\partial}{\partial m^{2}} + \gamma_{\xi}(u)\,\xi\frac{\partial}{\partial \xi}\Bigg] \Gamma^{(n)}(P_{1},...,P_{n};m,u,\mu) = 0,\hspace{1mm} 
\end{eqnarray}
where the $\beta$-function, field $\gamma$, mass $\gamma_{m^{2}}$ and non-minimal coupling constant $\gamma_{\xi}$ anomalous dimensions are defined by
\begin{eqnarray}
\beta(u) = \mu\frac{\partial u}{\partial\mu}\Bigg\vert_{B}, \gamma(u) = \frac{1}{2}\mu\frac{\partial}{\partial\mu}Z_{\phi}\bigg\vert_{B}, \gamma_{m^{2}}(u) = \frac{\mu}{m^{2}}\frac{\partial m^{2}}{\partial\mu}\Bigg\vert_{B}, \gamma_{\xi}(u) = \frac{\mu}{\xi}\frac{\partial \xi}{\partial\mu}\Bigg\vert_{B}.\nonumber \\
\end{eqnarray}
In the renormalized theory, the bare quantities are substituted by their renormalized counterparts. Thus, we have to employ, from now on, the renormalized internal line which corresponds to the renormalized free Green's function $G_{0}(q) = \parbox{12mm}{\includegraphics[scale=1.0]{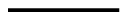}}$ expanded in normal coordinates and given by \cite{Bunch.Parker}
\begin{eqnarray}
G_{0}(q) = \frac{1}{q^{2} + m^{2}} + \frac{(1/3 - \xi)R}{(q^{2} + m^{2})^{2}} - \frac{2R_{\mu\nu}q^{\mu}q^{\nu}}{3(q^{2} + m^{2})^{3}}. 
\end{eqnarray}
Our considerations will be limited to the linear order in $R$ and $R_{\mu\nu}$. We could inspect if divergent terms proportional to $R^2$ and $R_{\mu\nu}^2$ would be generated at one-loop order. In fact, the free propagator to second order in $R$ and $R_{\mu\nu}$ \cite{Bunch.Parker} is given by two terms, namely the $(1/6 - \xi)^{2}R^{2}(q^{2} + m^{2})^{-3}$ + $a_{\mu\nu}(q^{2} + m^{2})^{-1}(\partial/\partial q_{\mu})(\partial/\partial q_{\nu})(q^{2} + m^{2})^{-1}$ ones, where $a_{\mu\nu}$ is a constant independent of momentum $q$ and written in terms of some second order functionals of $R$. From power counting arguments, the superficial degree of divergence assumes the negative value equal to $-2$. Thus we see that these two terms are convergent and do not produce any divergent contributions. It is noteworthy to mention that the problem approached in the present article is a semiclassical one, \emph{i. e.} the problem of a quantum field interacting with a curved classical background. Although it is not the purpose of the present work, quantum gravity effects could be considered by substituting the curved classical background by its quantum gravity counterpart thus giving rise to the need of including into the problem the graviton propagator in curved spacetime \cite{Galley.Hu} given by $D_{abcd}(q) = \frac{2P_{abcd}^{(2)}}{q^{2}} - \frac{P_{abcd}^{(0)}}{q^{2}} - \frac{R_{acbd} + R_{adbc}}{(q^{2})^{2}}  - \frac{1}{3}\frac{k^{s}k^{t}}{(q^{2})^{3}}\left[ \delta_{ac}R_{bsdt} + \delta_{ad}R_{bsct} + \delta_{bc}R_{asdt} + \delta_{ab}R_{csdt} + \frac{4}{(d - 2)^{2}}(\delta_{ab}R_{csdt} + \delta_{cd}R_{asbt})\right]$, where $P_{abcd}^{(2)}$ and $P_{abcd}^{(0)}$ are the spin-$2$ and spin-$0$ projection operators \cite{Hamber}, respectively. Some field theoretical renormalization group calculations were performed early \cite{I.JackandH.OsbornI,I.JackandH.OsbornII} in a standard procedure that involves both space and momentum coordinates. Our computations are performed at dimensions smaller than four up to next-to-leading order where we use only momentum coordinates. Now we have to apply the BPHZ method, up to next-to-leading order, to renormalize the theory. The evaluated expressions for the Feynman diagrams and counterterms are displayed just below
\begin{eqnarray}
\left(\parbox{12mm}{\includegraphics[scale=1.0]{fig6.eps}}\right)\Biggr|_{m^{2} = R = 0} =  -\frac{u^{2}P^2}{8\epsilon} \left[ 1 + \frac{1}{4}\epsilon -2\epsilon\, J_{3}(P^{2}) \right],
\end{eqnarray}
\begin{eqnarray}
\parbox{12mm}{\includegraphics[scale=1.0]{fig7.eps}}\bigg|_{m^{2} = R = R_{\mu\nu} = 0} = \frac{P^2u^{3}}{6\epsilon^{2}} \left[1 + \frac{1}{2}\epsilon - 3\epsilon\, J_{3}(P^{2})\right],
\end{eqnarray}
\begin{eqnarray}
\parbox{10mm}{\includegraphics[scale=1.0]{fig26.eps}} = -\frac{3P^{2}u^{3}}{16\epsilon^{2}}\Biggr[1 + \frac{1}{4}\epsilon - 2\epsilon\, J_{3}(P^{2}) \Biggr],
\end{eqnarray}
\begin{eqnarray}
&& \parbox{10mm}{\includegraphics[scale=1.0]{fig10.eps}} = \frac{\mu^{\epsilon}u^{2}}{\epsilon} \Biggr[1 - \frac{1}{2}\epsilon - \frac{1}{2}\epsilon J(P^{2}) - \frac{(\xi - 1/6)R}{2\mu^{2}}\epsilon J_{\xi R}(P^{2}) +  \frac{R}{6\mu^{2}}\epsilon J_{R}(P^{2}) - \nonumber \\ &&  \frac{R_{\mu\nu}P^{\mu}P^{\nu}}{3\mu^{4}}\epsilon J_{R_{\mu\nu}}(P^{2}) \Biggr],
\end{eqnarray}
\begin{eqnarray}
&& \parbox{16mm}{\includegraphics[scale=1.0]{fig11.eps}} = -\frac{\mu^{\epsilon}u^{3}}{\epsilon^{2}} \Biggr[1 - \epsilon - \epsilon J(P^{2}) - \frac{(\xi - 1/6)R}{\mu^{2}}\epsilon J_{\xi R}(P^{2}) + \frac{R}{3\mu^{2}}\epsilon J_{R}(P^{2}) - \nonumber \\ &&  2\frac{R_{\mu\nu}P^{\mu}P^{\nu}}{3\mu^{4}}\epsilon J_{R_{\mu\nu}}(P^{2}) \Biggr],
\end{eqnarray}
\begin{eqnarray}
&&\parbox{12mm}{\includegraphics[scale=0.8]{fig21.eps}} = -\frac{\mu^{\epsilon}u^{3}}{2\epsilon^{2}} \Biggr[1 - \frac{1}{2}\epsilon - \epsilon J(P^{2}) - \frac{(\xi - 1/6)R}{\mu^{2}}\epsilon J_{\xi R}(P^{2}) +  \frac{R}{3\mu^{2}}\epsilon J_{R}(P^{2}) - \nonumber \\ &&  2\frac{R_{\mu\nu}P^{\mu}P^{\nu}}{3\mu^{4}}\epsilon J_{R_{\mu\nu}}(P^{2})\Biggr],
\end{eqnarray}
\begin{eqnarray}
&&\parbox{12mm}{\includegraphics[scale=1.0]{fig13.eps}} =   \frac{\mu^{\epsilon}u^{3}}{2\epsilon} \Biggr[\frac{m^{2}}{\mu^{2}}J_{\xi R}(P^{2}) + \frac{(\xi - 1/6)R}{\mu^{2}}J_{\xi R}(P^{2}) -  \frac{(\xi - 1/6)Rm^{2}}{\mu^{4}}J_{4\xi R}(P^{2}) + \nonumber \\ &&  \frac{Rm^{2}}{9\mu^{4}}J_{4R}(P^{2}) - \frac{20R_{\mu\nu}P^{\mu}P^{\nu}m^{2}}{9\mu^{6}}J_{4R_{\mu\nu}}(P^{2})\Biggr],
\end{eqnarray}
\begin{eqnarray}
&&\parbox{10mm}{\includegraphics[scale=1.0]{fig25.eps}} = \frac{3\mu^{\epsilon}u^{3}}{2\epsilon^{2}} \Biggr[1 - \frac{1}{2}\epsilon - \frac{1}{2}\epsilon J(P^{2}) - \frac{(\xi - 1/6)R}{2\mu^{2}}\epsilon J_{\xi R}(P^{2}) +  \frac{R}{6\mu^{2}}\epsilon J_{R}(P^{2}) - \nonumber \\ &&  \frac{R_{\mu\nu}P^{\mu}P^{\nu}}{3\mu^{4}}\epsilon J_{R_{\mu\nu}}(P^{2}) \Biggr],
\end{eqnarray}
\begin{eqnarray}
&&\parbox{12mm}{\includegraphics[scale=1.0]{fig24.eps}} =  -\frac{\mu^{\epsilon}u^{3}}{4\epsilon} \Biggr[\frac{m^{2}}{\mu^{2}}J_{\xi R}(P^{2}) + \frac{(\xi - 1/6)R}{\mu^{2}}J_{\xi R}(P^{2}) - \frac{(\xi - 1/6)Rm^{2}}{\mu^{4}}J_{4\xi R}(P^{2}) + \nonumber \\ &&  \frac{Rm^{2}}{9\mu^{4}}J_{4R}(P^{2}) - \frac{20R_{\mu\nu}P^{\mu}P^{\nu}m^{2}}{9\mu^{6}}J_{4R_{\mu\nu}}(P^{2})\Biggr],
\end{eqnarray}
\begin{eqnarray}
\parbox{12mm}{\includegraphics[scale=1.0]{fig1.eps}}\Biggr|_{R = 0} =
\frac{m^{2}u}{\epsilon}\left[ 1 - \frac{1}{2}\epsilon\ln\left(\frac{m^{2}}{\mu^{2}}\right)\right],
\end{eqnarray}
\begin{eqnarray}
\parbox{10mm}{\includegraphics[scale=1.0]{fig2.eps}}\Biggr|_{R = 0} = - \frac{m^{2}u^{2}}{\epsilon^{2}}\left[ 1 - \frac{1}{2}\epsilon - \epsilon\ln\left(\frac{m^{2}}{\mu^{2}}\right)\right],
\end{eqnarray}
\begin{eqnarray}
\parbox{12mm}{\includegraphics[scale=1.0]{fig22.eps}}\Biggr|_{R = 0} =  \frac{m^{2}g^{2}}{2\epsilon^{2}}\left[ 1 - \frac{1}{2}\epsilon - \frac{1}{2} \epsilon\ln\left(\frac{m^{2}}{\mu^{2}}\right)\right],
\end{eqnarray}
\begin{eqnarray}
\parbox{12mm}{\includegraphics[scale=1.0]{fig23.eps}}\Biggr|_{R = 0} =  \frac{3m^{2}u^{2}}{2\epsilon^{2}}\left[ 1 - \frac{1}{2} \epsilon\ln\left(\frac{m^{2}}{\mu^{2}}\right)\right],
\end{eqnarray}
\begin{eqnarray}
\left(\parbox{12mm}{\includegraphics[scale=1.0]{fig6.eps}}\right)\Biggr|_{P^{2} = R = 0} =   -\frac{3m^{2}u^{2}}{2\epsilon^{2}}\left[ 1 + \frac{1}{2}\epsilon -\epsilon\ln\left(\frac{m^{2}}{\mu^{2}}\right)\right],
\end{eqnarray}
\begin{eqnarray}
\parbox{12mm}{\includegraphics[scale=1.0]{fig1.eps}}\Biggr|_{m^{2} = 0} =
\frac{(\xi - 1/6)Ru}{\epsilon}\left[ 1 - \frac{1}{2}\epsilon - \frac{1}{2}\epsilon\ln\left(\frac{m^{2}}{\mu^{2}}\right)\right] + \frac{Ru}{36},
\end{eqnarray}
\begin{eqnarray}
\parbox{10mm}{\includegraphics[scale=1.0]{fig2.eps}}\Biggr|_{m^{2} = 0} = - \frac{(\xi - 1/6)Ru^{2}}{\epsilon^{2}}\left[ 1 - \frac{3}{2}\epsilon - \epsilon\ln\left(\frac{m^{2}}{\mu^{2}}\right)\right] - \frac{Ru^{2}}{18\epsilon},
\end{eqnarray}
\begin{eqnarray}
\parbox{12mm}{\includegraphics[scale=1.0]{fig22.eps}}\Biggr|_{m^{2} = 0} =  \frac{(\xi - 1/6)Ru^{2}}{2\epsilon^{2}}\left[ 1 - \epsilon - \frac{1}{2} \epsilon\ln\left(\frac{m^{2}}{\mu^{2}}\right)\right] + \frac{Ru^{2}}{36\epsilon},
\end{eqnarray}
\begin{eqnarray}
\parbox{12mm}{\includegraphics[scale=1.0]{fig23.eps}}\Biggr|_{m^{2} = 0} =  \frac{3(\xi - 1/6)Ru^{2}}{2\epsilon^{2}}\left[ 1 - \frac{1}{2}\epsilon - \frac{1}{2} \epsilon\ln\left(\frac{m^{2}}{\mu^{2}}\right)\right],
\end{eqnarray}
\begin{eqnarray}
\left(\parbox{12mm}{\includegraphics[scale=1.0]{fig6.eps}}\right)\Biggr|_{P^{2} = m^{2} = 0} =   -\frac{3(\xi - 1/6)Ru^{2}}{2\epsilon^{2}}\left[ 1 - \frac{1}{2}\epsilon -\epsilon\ln\left(\frac{m^{2}}{\mu^{2}}\right)\right] + \frac{Ru^{2}}{48\epsilon},
\end{eqnarray}
where
\begin{eqnarray}\label{uhduhufgjg}
J(P^{2}) = \int_{0}^{1}dx \ln \left[\frac{x(1-x)P^{2} + m^{2}}{\mu^{2}}\right],
\end{eqnarray}
\begin{eqnarray}\label{uhduhufgjgdhg}
&& J_{3}(P^{2}) = \nonumber \\ && \int_{0}^{1}\int_{0}^{1}dxdy(1-y)\ln \Biggl\{\frac{y(1-y)P^{2}}{\mu^{2}} + \left[1-y + \frac{y}{x(1-x)}  \right]\frac{m^{2}}{\mu^{2}}\Biggr\},
\end{eqnarray}
\begin{eqnarray}\label{ugujjgdhg}
J_{\xi R}(P^{2}) =  \int_{0}^{1}d x\frac{1}{\frac{x(1 - x)P^{2}}{\mu^{2}} + \frac{m^{2}}{\mu^{2}}},
\end{eqnarray}
\begin{eqnarray}\label{ugujdfjgdhg}
J_{R}(P^{2}) =  \int_{0}^{1}d x\frac{x(1 - x)}{\frac{x(1 - x)P^{2}}{\mu^{2}} + \frac{m^{2}}{\mu^{2}}},
\end{eqnarray}
\begin{eqnarray}\label{ufgfghujjgdhg}
J_{R_{\mu\nu}}(P^{2}) =  \int_{0}^{1}d x\frac{x^{2}(1 - x)^{2}}{\left[\frac{x(1 - x)P^{2}}{\mu^{2}} + \frac{m^{2}}{\mu^{2}}\right]^{2}},
\end{eqnarray}
\begin{eqnarray}\label{ufgujjgdhg}
J_{4\xi R}(P^{2}) =  \int_{0}^{1}d x\frac{(1-x)}{\left[\frac{x(1 - x)P^{2}}{\mu^{2}} + \frac{m^{2}}{\mu^{2}}\right]^{2}},
\end{eqnarray}
\begin{eqnarray}\label{ufgujjgdhg}
J_{4R}(P^{2}) =  \int_{0}^{1}d x\frac{(1 - x)^{2}(5x + 1)}{\left[\frac{x(1 - x)P^{2}}{\mu^{2}} + \frac{m^{2}}{\mu^{2}}\right]^{2}},
\end{eqnarray}
\begin{eqnarray}\label{ufgujjgdhg}
J_{4R_{\mu\nu}}(P^{2}) =  \int_{0}^{1}d x\frac{x^{2}(1 - x)^{3}}{\left[\frac{x(1 - x)P^{2}}{\mu^{2}} + \frac{m^{2}}{\mu^{2}}\right]^{3}}.
\end{eqnarray}
We observe that the Feynman diagrams and counterterms are expressed in terms of the momentum-dependent integrals of Eqs. (\ref{uhduhufgjg})-(\ref{ufgujjgdhg}) and the parameters characterizing the curved spacetime, namely the $R$ and $R_{\mu\nu}$ ones. Then, by performing the computation of curved spacetime $\beta$-function and anomalous dimensions we have that these momentum-dependent integrals and the curved spacetime parameters $R$ and $R_{\mu\nu}$ are canceled out in the middle of the calculations. Thus, we obtain, at least up to the next-to-leading order approached here, that the curved spacetime $\beta$-function $\beta(u)$ and anomalous dimensions $\gamma_{\phi}(u)$ and $\gamma_{m^{2}}(u)$ are the same as their flat spacetime counterparts. Then the corresponding curved spacetime critical exponents values to be obtained must be the same as that of flat spacetime \cite{Wilson197475}.
\begin{eqnarray}\label{reewriretjgjk}
\beta(u) = -\epsilon u + \frac{N + 8}{6} u^{2} - \frac{3N + 14}{12}u^{3},
\end{eqnarray} 
\begin{eqnarray}\label{jkjkpfgjrftj}
\gamma_{\phi}(u) = \frac{N + 2}{72}u^{2} - \frac{(N + 2)(N + 8)}{1728}u^{3},
\end{eqnarray} 
\begin{eqnarray}\label{gfydsguyfsdgufa}
\gamma_{m^{2}}(u) = \frac{N + 2}{6}u - \frac{5(N + 2)}{72}u^{2}.
\end{eqnarray} 
\begin{eqnarray}\label{gfydsgdduyfsdgufa}
\gamma_{\xi}(u) = \frac{N + 2}{6}\Bigg(\xi - \frac{1}{6}\Bigg)u - \frac{5(N + 2)}{72}\Bigg(\xi - \frac{7}{30}\Bigg)u^{2}.\quad
\end{eqnarray}
The nontrivial solution of the curved spacetime $\beta$-function is used for evaluating the nontrivial radiative quantum corrections to the curved spacetime critical exponents and is given by 
\begin{eqnarray}\label{fghyagyaguhd}
u^{*} = \frac{6\epsilon}{(N + 8)}\left\{ 1 + \epsilon\left[ \frac{3(3N + 14)}{(N + 8)^{2}} \right]\right\}.
\end{eqnarray}
The curved spacetime critical exponents can be computed through the relations $\eta\equiv \gamma_{\phi}(u^{*})$ and $\nu^{-1}\equiv 2 - \gamma_{m^{2}}(u^{*})$. We then obtain their values up to next-to-leading level 
\begin{eqnarray}\label{eta}
\eta = \frac{(N + 2)\epsilon^{2}}{2(N + 8)^{2}}\left\{ 1 + \epsilon\left[ \frac{6(3N + 14)}{(N + 8)^{2}} -\frac{1}{4} \right]\right\},
\end{eqnarray}
\begin{eqnarray}\label{nu}
\nu = \frac{1}{2} + \frac{(N + 2)\epsilon}{4(N + 8)} +  \frac{(N + 2)(N^{2} + 23N + 60)\epsilon^{2}}{8(N + 8)^{3}}.\quad
\end{eqnarray}
The four remaining curved spacetime critical exponents can be computed through the four scaling relations \cite{Stanley}. This means that the diffeomorphism symmetry has not affected the critical exponents values. This fact is in agreement with the universality hypothesis, at least at the next-to-leading order considered here, \emph{i. e.} that the critical exponents values can be affected by a symmetry mechanism only if that mechanism is one present in the internal space of the field and not in the one in which the field is embedded, which is the case of the present work. Now we proceed to our conclusions.

\section{Conclusions}\label{Conclusions}

\par We have evaluated analytically the critical exponents for massive O($N$) $\lambda\phi^{4}$ scalar field theories in curved spacetime. The aim of this task was to probe the effect of diffeomorphism symmetry on the curved spacetime critical indices values. For that, we have employed the field-theoretic renormalization group approach in the BPHZ method for renormalizing the theory. We have obtained the critical indices up to next-to-leading loop order and have shown that their values are independent of the curved spacetime parameters and thus are the same as their flat spacetime counterparts, at least up to that loop level. This result shows that the critical exponents values are insensible to the diffeomorphism symmetry, since this symmetry is present in the spacetime where the field is defined and not in its internal one. Thus the universality hypothesis remains intact, at least up to the loop level and the linear approximation in the curved spacetime parameters considered here. Furthermore, the present work opens a new research branch, that of considering possible effects of the diffeomorphism symmetry on the critical properties of systems undergoing a continuous phase transition as corrections to scaling and finite-size scaling for bulk critical exponents and critical exponents in geometries subjected to different boundary conditions as well as for amplitude ratios.

\section*{Acknowledgments}
HASC would like to thank CAPES (brazilian funding agency) for financial support.

\bibliography{apstemplate}

\end{document}